\documentstyle{l-aa}
\input{epsf}
\begin{document}
\thesaurus{02.04.01, 08.14.1, 08.16.6, 02.07.01}
\title{Radial pulsations and stability of protoneutron stars}
\author{D. Gondek
\and P. Haensel
\and J.L. Zdunik}

\institute{N. Copernicus Astronomical Center, Polish
           Academy of Sciences, Bartycka 18, PL-00-716 Warszawa, Poland\\
{\em e-mail: dorota, haensel, jlz@camk.edu.pl}}
\offprints{D. Gondek}
\date{}
\maketitle
%\markboth{D. Gondek et al.: Pulsations and stability of protoneutron stars}
%
%==========================================================
\begin{abstract}
Radial pulsations of newborn neutron stars (protoneutron stars) are
studied for a range of internal temperatures and entropies 
per baryon predicted by the existing numerical simulations. 
Protoneutron star models are constructed using a realistic  
equation of state of hot dense matter, and under various
assumptions concerning stellar interior (large trapped lepton
number, zero trapped lepton number,  isentropic,  isothermal). 
Under prevailing conditions, linear oscillations of a protoneutron
star can be treated as adiabatic, and evolutionary effects can
be neglected on dynamic timescale. The dynamic behavior is
governed by the adiabatic index, which in turn depends on the
physical state of the stellar interior. The eigenfrequencies 
of the lowest radial modes of linear, adiabatic pulsations are
calculated. Stability  of protoneutron stars
 with respect to small radial perturbations is studied, and 
 the validity of the  static stability criteria is discussed. 
\keywords{dense matter -- stars: neutron -- stars: pulsars}

\end{abstract}
%

%%%%%%%%%%%%%%%%%%%%%%%%%%%%%%%%%%
\section{Introduction }
 Newly born neutron stars are hot and
lepton rich objects, quite different from ordinary low
temperature, lepton poor neutron stars. In view of these
differences, newly born neutron stars  are called 
{\it protoneutron} stars; they transform into standard neutron
stars on a timescale of the order of ten seconds, needed for the 
loss of a significant lepton number excess via emission of
neutrinos trapped in dense, hot interior. 

In view of the fact that the typical evolution timescale of
a protoneutron star (seconds) is some three orders of magnitude
longer, than the dynamical timescale for this objects (milliseconds), 
one can study  its evolution in the quasistatic approximation
(Burrows \& Lattimer 1986). Static properties  of
protoneutron stars, under various assumptions concerning
composition and equation of state (EOS) of hot, dense stellar interior
 were studied by numerous authors (Burrows \& Lattimer 1986,
Takatsuka 1995, Bombaci et al. 1995, Bombaci 1996, 
Prakash et al. 1997). 

The scenario of transformation of a protoneutron star  into a
neutron star could be strongly influenced by a phase transition
in the central region of the star. Brown and Bethe (1994)
suggested a phase transition implied by the $K^-$ condensation
at supranuclear densities. Such a $K^-$ condensation could
dramatically soften the equation of state of dense matter,
leading to a low maximum allowable mass of neutron stars. 
In such a case, the massive protoneutron stars could be
stabilized by the effects of high temperature and of the
presence of trapped neutrinos, and this would lead to maximum
baryon mass of  protoneutron stars larger by some $0.2~M_\odot$ than
that of cold neutron stars. The deleptonization and cooling of
 protoneutron stars of baryon mass exceeding the maximum
allowable baryon mass for neutron stars,   would  then inevitably 
lead to their collapse  into black holes. The dynamics of 
such a process was
recently studied by Baumgarte et al. (1996). It should be
mentioned, however, that the very possibility of existence of
the kaon condensate  (or other exotic phases of matter, such as
the pion condensate, or the quark matter) at neutron star
densities is far from being  established. Recently, for
instance, 
Pandharipande et al. (1995) pointed out, that kaon-nucleon and
nucleon-nucleon correlations in dense matter raise
significantly the threshold density for kaon condensation.
 In view of these uncertainties, we will
restrict in the present paper to a standard model of dense
matter, composed of nucleons and leptons. 

The calculations of the static models of protoneutron stars should be
considered as a first step in the studies of these objects. It
is clear, in view of the dynamical scenario of their formation,
that protoneutron stars are far from being static. 
 The formation scenario
involves  compression (with overshoot of central density) and a 
hydrodynamical bounce, so that a newborn protoneutron star
begins its life in a highly excited state, pulsating around its
quasistatic equilibrium. If the rotation of protoneutron star is
not too rapid, the coupling of radial and non-radial pulsations
is weak. Such a  scenario of formation leads to a 
preferential excitation of the radial modes. Having constructed
the static model of a protoneutron star, one should thus answer the
question about the stability of the static configuration with
respect to radial perturbations (standard stability criteria are
valid only for cold neutron stars). Clearly, both high temperature 
 and large trapped lepton number will influence the spectrum of
radial eigenfrequencies of protoneutron stars, implying
differences with respect to the case of cold neutron stars.

In the present paper we study the radial pulsations of
protoneutron stars and their stability. Our models of
protoneutron stars are composed of a hot, neutrino-opaque
interior (hereafter referred to as ``hot interior''), separated
from much colder, neutrino-transparent 
envelope by a neutrinosphere. We will consider two limiting cases of the 
thermal state of the hot interior: isentropic, with entropy
per baryon $s=const.$, and  isothermal, with
$T_\infty=(g_{00})^{1/2}T=const.$ 
 ($T_\infty$ is the value of the temperature, measured by an 
observer at infinity, while $T$ is the value of the temperature 
 measured by a local observer). 

The first case, characteristic of a very
initial state of a protoneutron star, will simultaneously 
correspond to a significant trapped lepton number. The second
case corresponds to situation after the deleptonization of a
protoneutron star.  The  position of the neutrinosphere will 
be located using  a simple prescription
based on specific properties of  the neutrino opacity of hot
dense matter.
 In all cases, the equation of state of hot dense 
matter will be determined using one of the realistic models of
Lattimer and Swesty (1991).

The plan of the paper is as follows.  
 In Section 2 we describe
the physical state of the interior of protoneutron star, with
particular emphasis on the EOS of the hot interior at various
stages of evolution of a protoneutron star. We explain also our
prescription for locating the 
neutrinosphere of a protoneutron star, and we  
give  some details concerning the assumed temperature profile within
protoneutron star. 
 Equations of state, corresponding to different physical
situations, are described in Section 3. We also present there
static models of protoneutron stars, calculated for various
assumptions concerning the hot stellar interior. In Section 4 we
compare various timescales, characteristic of evolution and
dynamics of a protoneutron star, which are essential for
approximations used in the treatment of pulsations. 
 Section 5 is devoted to the formulation of the problem of
linear, adiabatic, radial pulsations of protoneutron stars. 
Both pulsations and stability involve the adiabatic index of
pulsating stellar interior; our results for this important
quantity are presented in Section 6. Numerical results for the
eigenfrequencies of the lowest modes of radial pulsations  and
the problem of stability of protoneutron stars are presented in
Section 7.  Finally, Section 8 contains a discussion of our
results and conclusion. 
%
%%%%%%%%%%%%%%%%%%%%%%%%%%%%%%%%%%%%%%%%%%%%%%%%%%%%%%%%%
\section{Physical model of the interior of protoneutron stars}
We will consider a protoneutron star (PNS) 
just after its formation. We will assume it has a
well defined neutrinosphere, which separates a hot,
neutrino-opaque interior from colder, neutrino-transparent outer
 envelope. The  parameters, which determine the local state
of the matter in the hot interior are: baryon
(nucleon) number 
density $n$,  net electron fraction $Y_e
= (n_{e^-}-n_{e^+})/n$, and the net electron-neutrino
fraction $Y_{\nu}=Y_{\nu_e}-Y_{\bar\nu_e}$. The calculation of
the composition of hot matter and of its EOS is  described below.
%%%%%%%%%%%%%%%%%%%%%%%%
\subsection{Neutrino opaque core with nonzero trapped lepton number}
The situation described in this subsection 
is characteristic of
 the very initial stage of existence
of a PNS. Matter is assumed  to be 
composed of nucleons (both free and bound in
nuclei) and leptons (electrons and neutrinos; for the sake of 
simplicity, we
do not include muons). All constituents
of the matter (plus photons) are in thermodynamic equilibrium at
given values of $n$, $T$ and $Y_l=Y_e+Y_\nu$. The
composition of the matter is calculated from the condition of
 beta equilibrium, combined with the condition of a fixed
$Y_l$, 
%%%%%%%%%
\begin{eqnarray}
\mu_p + \mu_e &=& \mu_n + \mu_{\nu_e}~,\nonumber \\
Y_l&=&Y_e+Y_\nu~,
\label{mu.trapL}
\end{eqnarray}
%%%%%%%%%%%%%%%
where $\mu_{\rm j}$ are the chemical potentials of matter
constituents. At the very initial stage we expect $Y_l\simeq
0.4$. Electron neutrinos are degenerate, with $\mu_{\nu_e}\gg T$
(in what follows we measure $T$ in energy units). The
deleptonization, implying the decrease of $Y_l$, 
 occurs due to diffusion of neutrinos outward
(driven by the $\mu_{\nu_e}$ gradient), on a timescale of 
seconds (Sawyer \& Soni 1979, Bombaci et al. 1996). 
The diffusion of highly degenerate neutrinos from the central
core is a dissipative
process, resulting in a significant {\it heating} of the
neutrino-opaque core (Burrows \& Lattimer 1986). 
%%%%%%%%%%%%%%%%%%%%%%%%
\subsection{Neutrino opaque core with $Y_\nu = 0$}
This is the limiting case, reached after complete deleptonization. 
There is no trapped lepton number, so that $Y_l=Y_e$ and
$Y_{\nu_e}=Y_{\bar\nu_e}$, and therefore 
$\mu_{\nu_e}=\mu_{\bar\nu_e}=0$. Neutrinos trapped within the
hot interior do not therefore influence the beta equilibrium of nucleons,
electrons and positrons, and  for given $n$ and $T$ the
equilibrium value of $Y_e$ is determined from
%%%%%%%%%
\begin{eqnarray}
\mu_p + \mu_e &=& \mu_n~,
\label{mu.free}
\end{eqnarray}
%%%%%%%%%%%%%%%
while $\mu_{e^+}=-\mu_e$. In practice, this approximation can
be used  as soon as electron neutrinos become non-degenerate
within the opaque core, 
 $\mu_{\nu_e}< T$, which occurs after  some $\sim 10$ seconds
(Sawyer \& Soni 1979, Prakash et al. 1997). 
%%%%%%%%%%%%%%%%%%%%%%%%%
\subsection{Neutrinosphere  and  the temperature profile}
In principle, the temperature (or entropy per nucleon) profile
within a PNS has to be determined via evolutionary calculation,
starting from some initial state, and taking into account
relevant transport processes in the PNS interior, as well as
neutrino emission from PNS.  Transport
processes within neutrino-opaque interior occur on timescales of
seconds, some three orders of magnitude longer than dynamical
timescales. Convection can shorten neutrino transport timescale, 
 but still the time needed for the deleptonization of the 
neutrino opaque core is then much longer than the dynamical 
timescale. The very outer layer of a PNS becomes rapidly
transparent to neutrinos, deleptonizes, and cools on a very
short timescale 
via  $e^-e^+$ pair annihilation and plasmon decay to $T<1$ MeV.
It seemed thus natural to model the thermal structure of the PNS
interior by a hot core limited by a neutrinosphere, and a 
much cooler,  neutrino transparent outer envelope. 
 The transition through the neutrinosphere
is accompanied by a temperature drop, which takes place over
some interval of density just above the ``edge'' of the hot
neutrino-opaque core, situated at some $n_\nu$. 

In view of the uncertainties in the actual temperature profiles
within the hot interior of PNS, we considered two extremal
situations for $n>n_\nu$, 
corresponding to an isentropic and an isothermal hot interior. In the
first case, hot interior was characterized by a  constant entropy
per baryon $s=const.$. In the case of trapped lepton number, 
this leads to the EOS of the type: pressure 
$P=P(n,~[s,Y_l])$, energy density 
divided by $c^2$ (energy-mass density), 
$\rho=\rho(n,~[s,Y_l])$, and temperature 
$T=T(n,~[s,Y_l])$, 
 with  fixed  $s$ and $Y_l$. 
 This EOS  will be denoted by EOS[$s,Y_l$]. 

The condition of isothermality, which in the static case
corresponds to a vanishing heat flux,  is more
complicated. Due to the curvature of the space-time within PNS, 
the condition of isothermality (thermal equilibration) 
corresponds to 
the constancy of $T_\infty=(g_{00})^{1/2}T$ 
(see, e.g., Zeldovich \& Novikov 1971, chapter 10.6). 
 Actually, the isothermal state
within  the hot interior will be reached on a timescale
 corresponding to thermal equilibration, which is much longer
than the lifetime of a PNS. 
Nevertheless, we considered the
$T_\infty=const.$ models  because, as a limiting case
so different from the $s=const.$ one, it enables us to check the
dependence of our results for  PNS on the
thermal state of the hot interior. 

To determine the isothermal temperature profile in the hot interior  
after deleptonization,  
we use the condition of thermal equilibrium, 
 given  
by the constancy of the function $T_\infty  \equiv T(r)
 e^{{1\over2}\nu (r)}$ (where $\nu(r)$ is the metric fuction, 
 see Sect. 5). The relativistic condition of the 
isothermality can be rewritten as:
%%%%%%%%%%%%%%%%%%%%%%%%%%%%%
\begin{equation}
{{\rm d} \ln T \over {\rm d}r} =
{1\over \rho c^2 + P}{{\rm d}P\over {\rm d}r}~. 
\label{eq:isot}
\end{equation}
%%%%%%%%%%%%%%%%%%%%%%%%%%%%%
This formula enables us to determine the $T(n)$ profile,  
for given EOS,  
 {\it independently} of the specific structure of the stellar 
model under consideration.

Treating $n$, $T$  as thermodynamic variables for our equation of
state,  we can rewrite Eq. (\ref{eq:isot}) in the form:
%%%%%%%%%%%%%%%%%%%%%%%%%%%%%%
\begin{eqnarray}
{{\rm d} \ln T\over {\rm d} \ln n} 
 &=&{P\over \rho c^2 + P}\nonumber\\ 
 &\times&\left({\partial \ln P \over \partial \ln n}
\right) 
 \left\{1- {P\over \rho c^2 + P} 
\left({\partial \ln P \over \partial \ln T}\right) 
\right\}^{-1}~.
\label{eq:isot1}
\end{eqnarray}
%%%%%%%%%%%%%%%%%%%%%%%%%%%%%%%

Using the above formula we can construct  a specific isothermal 
 EOS,  describing hot isothermal interiors, and 
 parametrized by the boundary condition at the edge of 
the hot isothermal core - 
 the value of $T$ just below the neutrinosphere 
will be denoted by $T_{\rm b}$. 
 Thus in our relativistic calculations the set of the 
 ``isothermal'' stellar
configurations corresponds to stars with given $T_{\rm b}$. 
 Starting from $T=T_{\rm b}$, the  temperature increases inward, 
reaching its maximum value in the center of the star where 
$T=T_{\rm centr}$.
This central temperature is the function of the central 
density $\rho_{\rm centr}$
and is larger for a star with larger $\rho_{\rm centr}$.

Our calculation of the neutrinosphere within the hot PNS
interior is explained below. 
  For a given static PNS model, the neutrinosphere 
radius, $R_\nu$, has been
located through the condition
%%%%%%%%%
\begin{equation}
\int_{R_\nu}^R
{1\over \lambda_\nu(E_\nu)}
{\rm d}r_{\rm prop}= 1~,
\label{R_nu}
\end{equation}
%%%%%%%%%%%%%%%
where $\lambda_\nu$ is calculated at  the matter temperature, 
  $E_\nu$ is the mean energy of non-degenerate neutrinos at and 
  above the 
neutrinosphere, $E_\nu=3.15T_\nu$, and  $r_{\rm prop}$ is the
proper distance from the star center. 
We assumed that neutrino opacity above $R_\nu$
is dominated by the elastic scattering off nuclei and nucleons,
so that $\lambda_\nu = \lambda^0_\nu(n,T)/E_\nu^2$. Then,
we determined the value of the density at the neutrinosphere, 
 $n_\nu$,  for a given static PNS model, combining  Eq. 
 (\ref{R_nu})
with that of hydrostatic equilibrium, and readjusting
 accordingly the temperature profile  within the outer layers of
PNS.

Neutrino opacity in the outer layers of PNS can be well
approximated by $1/\lambda_\nu\simeq \kappa_0
{E_\nu}^2$. 
 Within a reasonable approximation 
$\kappa_0\simeq \gamma\rho $, 
where $\rho$ is the matter
density, and $\gamma=6\cdot 10^{-20}~{\rm
cm^{-1}}$ ($\rho$ in ${\rm g~cm^{-3}}$ and $E_\nu$ in MeV).  
The proper distance
near the neutrinosphere is given by ${\rm d}r_{\rm
prop}= (g_{rr})^{1/2}{\rm d}r
\simeq (1-2GM/Rc^2)^{-1/2}{\rm d}r$.
Using the definition of the neutrinosphere radius, $R_\nu$, we
can express the mass of the envelope above $R_\nu$ and the 
pressure at $R_\nu$, denoted by $P_\nu$,  in terms of
${E_\nu}=3.15{T_\nu}$. This leads to 
%%%%%%%%%%%%%%%%%%%
\begin{equation}
P_\nu =
2.23~10^{32}
~\widetilde{M}\widetilde{R}^{-2}
\left
(1-0.295{\widetilde{M}\over\widetilde{R}}\right)^{-{1\over 2}}
~{T_\nu}^{-2}~{\rm erg\over cm^3}~. 
\label{P_nu}
\end{equation}
%%%%%%%%%%%%%%%%%%%%%%%%%
where $\widetilde{M}=M/M_\odot$ and $\widetilde{R}=R/10$~km.

For a given stellar model, this approximate relation, combined
with Eq.(3),  enables one to determine, in a
self-consistent way, the values of $T,~n$, and $\rho$ at the
neutrinosphere. In practice, this was done assuming a specific
functional form of the temperature drop within the
neutrinosphere (a combination of Fermi functions), which yielded
a smooth transition between hot interior and cool envelope. In all cases,
the temperature profile was adjusted in such a way, that
neutrinosphere  was found within the region of the temperature
drop.  For a $1.4~M_\odot$ PNS, and an isothermal hot core with 
$T_{\rm b}=
15~$MeV, we found 
$T_\nu=4.3~$MeV, 
$n_\nu=2.0~10^{-3}~{\rm fm^{-3}}$, 
and $\rho_\nu=3.5~10^{12}~{\rm g~cm^{-3}}$. 
In the case of a very massive PNS with $M=2~M_\odot$ we
obtained  (for the same value of $T_{\rm b}$) 
$T_\nu=5.2~$MeV, 
$n_\nu=2.4~10^{-3}~{\rm fm^{-3}}$, 
and $\rho_\nu=4.1~10^{12}~{\rm g~cm^{-3}}$. These results are
in a reasonable agreement with those obtained in detailed numerical
simulations (see, e.g., Burrows \& Lattimer 1986). 
%%%%%%%%%%%%%%%%%%%%%
\section{Equation of state and static models of protoneutron stars}

The starting point for the construction of our EOS for the PNS
models was the model of hot dense matter of Lattimer and Swesty
(1991), hereafter referred to as LS. Actually, we used one
specific LS model, corresponding to the incompressibility
modulus at the saturation density of symmetric nuclear matter
$K=220~$MeV (this model will be hereafter referred to 
as LS-220). For $n>n_\nu$ we supplemented the LS-220 model
with contributions 
resulting from the presence of trapped neutrinos of three
flavours (electronic, muonic and tauonic) and of the 
corresponding antineutrinos. 

%
%
%%%%%%%%%%%%%%%%%
\begin{figure}     %1
\begin{center}
\leavevmode
\epsfxsize=8.5cm %\epsfysize=8.5cm 
\epsfbox{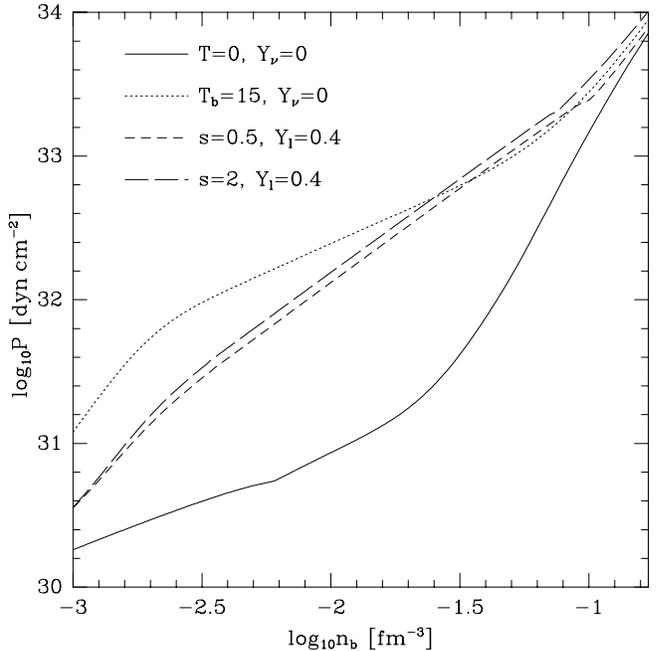}
\end{center}
\caption[]{
Pressure versus baryon density for our model of dense hot matter 
(based on the LS-220 model of the nucleon component of the 
matter), 
under various physical conditions, 
for the subnuclear densities $n<n_0=0.16~{\rm fm^{-3}}$.  
The curve $T=0$ corresponds to cold catalyzed matter.  
The curve corresponding to $s=0.5, Y_l=0.4$ is unphysical, but 
has been added in order to  visualize the importance of trapped lepton 
number at subnuclear densities.  The low density edge of the
hot, neutrino opaque core corresponds to  
 $n_\nu=2\times 10^{-3}~{\rm fm^{-3}}$.
}
\end{figure}
%%%%%%%%%%%%%%%%%
%%%%%%%%%%%%%%%%%
\begin{figure}     %2
\begin{center}
\leavevmode
\epsfxsize=8.5cm %\epsfysize=8.5cm 
\epsfbox{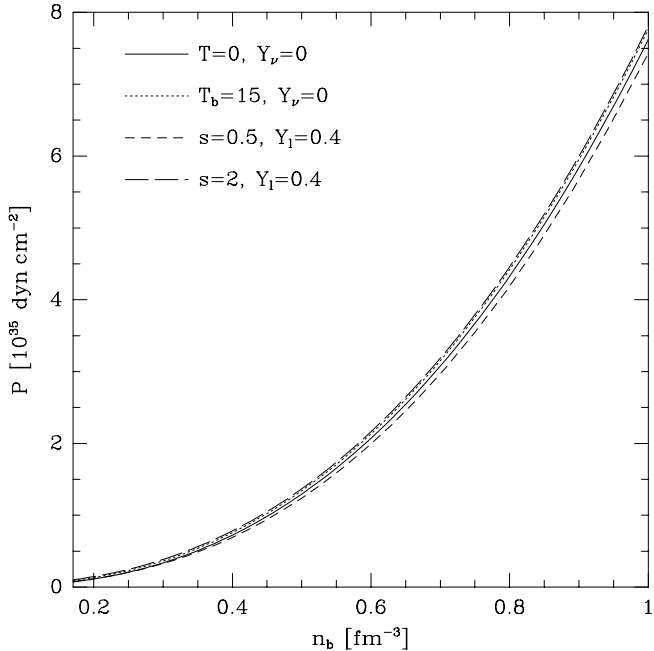}
\end{center}
\caption[]{
Pressure versus baryon density for our model of dense hot matter, 
under various physical conditions, for the  
supranuclear densities. 
The curve $T=0$ corresponds to cold catalyzed matter.  
The curve corresponding to $s=0.5, Y_l=0.4$ is unphysical, but 
has been added in order to  visualize the importance of trapped lepton 
number at subnuclear densities.  The low density edge of the
hot, neutrino opaque core corresponds to 
 $n_\nu=2\times 10^{-3}~{\rm fm^{-3}}$.
}
\end{figure}
%%%%%%%%%%%%%%%%%

In Figs. 1-3 we show our EOS in several cases, corresponding to
various physical condition in the hot, neutrino-opaque interior
of PNS. For the sake of comparison, we have also shown the EOS
for cold catalyzed matter, used for the calculation of the
(cold) NS models. 
In Fig. 1 we show our  EOS at subnuclear
densities.
At such densities, both the temperature and the
presence of trapped  neutrinos 
 lead to a significant increase of pressure,
 as compared to the cold catalyzed matter.
The constant $T_\infty$
 EOS stiffens considerably at lower
densities, which is due to the  weak dependence of
the thermal contribution (photons, neutrinos) on the baryon
density of the matter. 
  It is quite obvious, that $T_\infty=const.$ EOS 
becomes dominated by thermal effects for $n<10^{-2}~{\rm
fm^{-3}}$.  On the contrary, for the isentropic EOS, the effect of
the trapped 
lepton number ($Y_l=0.4$) turns out to be more important
than the thermal effects. This can be seen in Fig. 1, 
by comparing long-dashed curve, 
$[s=2,~Y_l=0.4]$, with a short-dashed line, which
corresponds to an artificial (unphysical) case with small
thermal effects, $[s=0.5,~Y_l=0.4]$. 

It is clear, that the correct location of the 
neutrinosphere, which separates hot interior from the colder
outer envelope, should be important for the determination of the
radius of PNS. 

Our EOS above nuclear density is plotted in Fig. 2. 
The presence of a trapped lepton number softens the EOS, while
thermal effects always
 lead to pressure increase  
with respect to that for cold catalyzed matter. 
The softening of the supranuclear EOS at fixed
$Y_l$ is due to 
the fact, that a significant trapped lepton number increases the
proton fraction, which implies the softening of the nucleon
contribution to the EOS.
  This is the reason why pressure for $[s=0.5,~Y_l=0.4]$ model
is smaller than in the case of cold, catalyzed matter.

%%%%%%%%%%%%%%%%%
\begin{figure}     %3
\begin{center}
\leavevmode
\epsfxsize=8.5cm \epsfbox{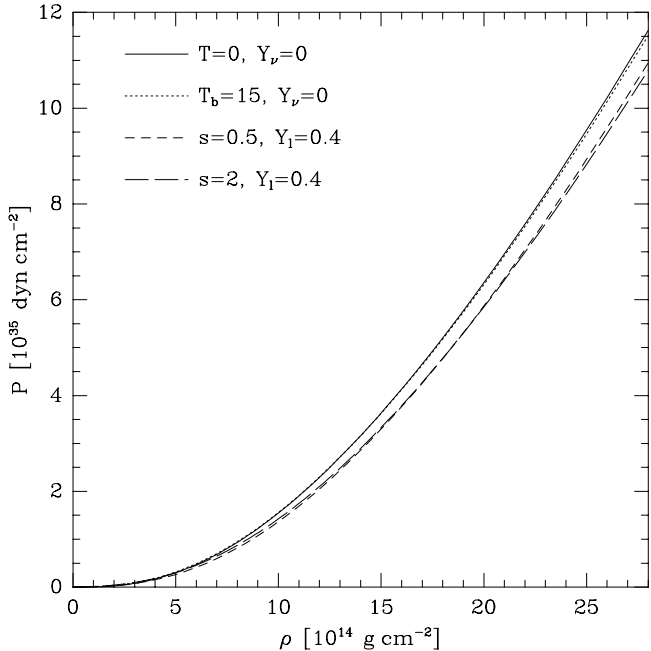}
\end{center}
\caption[]{
Pressure versus mass--energy density
for our model of dense hot matter, 
under various physical conditions.
The curve $T=0$ corresponds to cold catalyzed matter. 
}
\label{fig3}
\end{figure}
%%%%%%%%%%%%%%%%%%%%%%%%%%%%%%%%%%%

In the calculations of the stellar structure the 
 relevant EOS is of the form 
 $P=P(\rho)$,  because only pressure and mass--energy density
 appear in  the general relativistic equations of 
hydrostatic equilibrium.
It is obvious that thermal contribution to $\rho$ is always positive.
 Also, the contribution of trapped neutrinos (which do not 
contribute to baryon number density $n$) to $\rho$ is 
positive. However, large trapped lepton number implies an 
increased proton  fraction, which in turn implies a {\it softening} 
of the nucleon contribution to the pressure. The interplay of 
these effects leads to a characteristic {\it softening} of the 
$P(\rho)$ EOS for large trapped lepton number, visualized in 
Fig. 3.
  
It should be stressed, that in contrast to 
Bombaci et al. (1995) and Prakash et al. (1997) we used  a 
unified dense matter model, valid for both 
supranuclear and subnuclear densities. Also, the fact that we
use various assumptions about the $T$ and $s$ profiles within
PNS, enables us to study the relative importance of the 
temperature profile
and that of a trapped lepton number, for the PNS models. 

%%%%%%%%%%%%%%%%%
\begin{figure}     %4
\begin{center}
\leavevmode
\epsfxsize=8.5cm \epsfbox{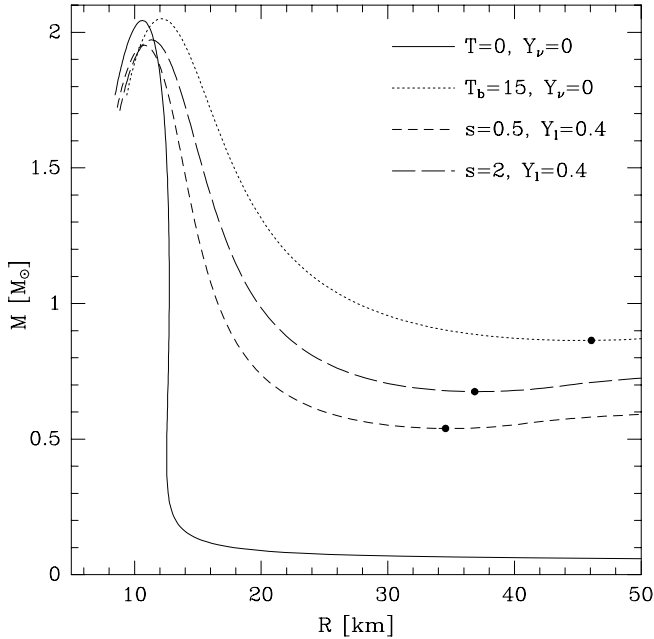}
\end{center}
\caption[]{
The gravitational mass versus stellar radius for
static models  
of the protoneutron stars and neutron stars, under various assumptions 
concerning the physical conditions within the stellar interior. 
The curve corresponding to $s=0.5, Y_l=0.4$ is unphysical, but has been 
added in order to visualize the relative   importance of the trapped lepton 
number and thermal effects. The curve $T=0$ corresponds
to cold neutron stars. 
}
\label{fig4}
\end{figure}
%%%%%%%%%%%%%%%%%%%%%%%%

%%%%%%%%%%%%%%%%%
\begin{figure}     %5
\begin{center}
\leavevmode
\epsfxsize=8.5cm \epsfbox{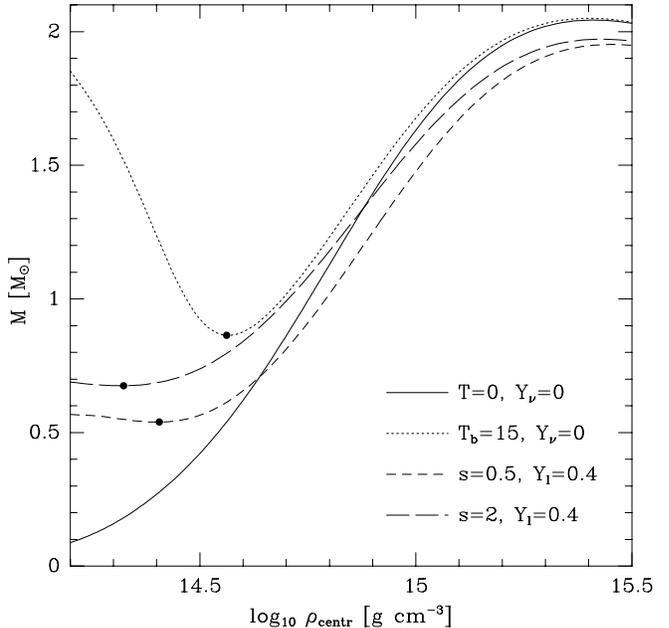}
\end{center}
\caption[]{
The gravitational mass versus central density for
static models  
of the protoneutron stars and neutron stars, under various assumptions 
concerning the physical conditions within the stellar interior. 
The curve corresponding to $s=0.5, Y_l=0.4$ is unphysical, but has been 
added in order to visualize the relative   importance of the trapped lepton 
number and thermal effects. The curve $T=0$ corresponds
to cold neutron stars. 
}
\label{fig5}
\end{figure}
%%%%%%%%%%%%%%%%%%%%%%%%

The mass-radius relation for the PNS models calculated using
various versions of our EOS for the hot interior is shown in
Fig. 4.  We assumed $n_\nu = 2\times 10^{-3}~{\rm fm^{-3}}$,
which was consistent with our definition of the neutrinosphere. 
For the sake 
of comparison, we also show the mass-radius relation for the
$T=0$ (cold catalyzed matter) EOS, which corresponds to cold 
neutron star models. In the case of an isothermal hot interior
with $T_{\rm b}=15$~MeV  we note a very small increase of
the maximum mass, as compared to the $T=0$ case (this result 
is consistent with those obtained by Bombaci
et al. 1995, Prakash et al. 1997, for models composed of nucleons 
and leptons only). However, the
effect on the mass-radius relation is quite strong, and
increases rapidly with decreasing stellar mass. In the case of
the isentropic EOS with a trapped lepton number, [$s=2,Y_l=0.4$],
the softening of the high-density EOS due to the trapped $Y_l$
leads 
to the decrease of $M_{\rm max}$ compared to the $T=0$ case; as far
as the value of $M_{\rm max}$ is concerned, the softening
effect of $Y_l$ prevails over that of finite $s$ (this is
consistent with results of Takatsuka 1995, Bombaci et al.
1995, and Prakash et al. 1997 for purely nucleonic models). However, the 
thermal effect on the stellar radius is 
important even in the case of  $Y_l=0.4$. 
This can be seen by comparing $[s=2,~Y_l=0.4]$ curve with that
corresponding to the unphysical, fictitious case of
$[s=0.5,~Y_l=0.4]$.
Let us notice, that 
for our EOS, 
configurations with maximum (minimum) gravitational mass 
 are simultaneously those with the maximum (minimum) baryon mass. 
The values of the minimum and maximum gravitational and baryon 
masses for various EOS of the PNS interiors, based on 
the LS-220 models of the nucleon component of dense matter, 
 are given in Table 1. The baryon (rest) mass of a PNS model 
is defined by $M_{\rm bar}=Am_0$, where $A$ is the total baryon 
number of a PNS, and $m_0$ is the mass of the hydrogen atom. 
%%%%%%%%%%%%%%%%%%%%%%%%%%%%%%%%%%%
\begin{table}
\caption[]{
Minimum and maximum gravitational and baryon masses of cold neutron star
($T=0$) and isothermal ($T_{\rm b}=15~{\rm MeV}, Y_\nu=0$) and isentropic
($s=2, Y_l=0.4$) protoneutron stars. Equations of state are based 
on the LS-220
model of the nucleon  component of the matter.
}
\centering
\begin{tabular}{lllll}
\hline
 & & & &\\
EOS  & $M_{\rm min}$ & $M_{\rm bar,min}$ 
& $M_{\rm max}$ & $M_{\rm bar,max}$ \\
 & $[M_{\odot}]$ &  $[M_{\odot}]$ & $[M_{\odot}]$ & $[M_{\odot}]$ \\
 & & & &\\     
\hline
 & & &\\
$T = 0$, $Y_\nu = 0$  & 0.054 & 0.055 & 2.044 & 2.406 \\
$T = 15$, $Y_\nu = 0$ & 0.864 & 0.892 & 2.050 & 2.391 \\
$s = 2$, $Y_l = 0.4$ & 0.675 & 0.676 & 1.972 & 2.183 \\
 & & &\\
\hline
\end{tabular}
\end{table}
\vspace{0.5cm}
%%%%%%%%%%%%%%%%%%%%%%% 

 We  find that the value of
$M_{\rm bar, max}$ is the largest one for the $T=0$ (cold catalyzed matter) 
EOS. For an isothermal hot star with $T_{\rm b}=15$~MeV  the total baryon mass 
 is slightly lower than $M_{\rm bar,max}^{[T=0]}$, 
 but for isentropic stars we see a significant
decrease of $M_{\rm bar, max}$ as compared to the $T=0$ case. 
A newborn hot protoneutron star with
the maximum gravitational mass $M_{\rm max}^{[s=2,~Y_l=0.4]}=1.97\,M_\odot$ 
 transforms, due to deleptonization and cooling, into a cold NS of 
gravitational mass of $1.9~M_\odot$, some  $0.15~M_\odot$ less than the maximum
gravitational mass of cold neutron stars (c.f. the analysis of Bombaci 1996).

For a given model of the neutrino opaque PNS core, the value of $M_{\rm min}$
depends on the location of the neutrinosphere.
However, this dependence is relatively weak. 
For example in the case of EOS $[s=2,~Y_l=0.4]$ a rather drastic variation of 
$n_\nu$  from $2\times 10^{-4}~{\rm fm^{-3}}$ to
$6\times 10^{-3}~{\rm fm^{-3}}$ implies change in 
$M_{\rm min}^{[s=2,~Y_l=0.4]}$ from  $0.686~M_\odot$ to  $0.649~M_\odot$,
respectively (the radius of these minimum--mass stars varies then 
from $44~{\rm km}$ to $32.8~{\rm km}$, respectively).

At a  given mass, the radius of a PNS is significantly larger
than that of a cold NS. 
 It should be stressed, however, that the value of radius
 turns out to be 
 quite sensitive to the location of the edge of
the hot neutrino-opaque interior (i.e., to the value of
$n_\nu$), 
especially for PNS  which are  not close to the $M_{\rm max}$ 
configuration. 
The choice of Prakash et al. (1997) would lead to a
much smaller effect on $R$. 

The role of the thermal and the lepton number effects are 
particularly pronounced in the mass - central density plots for the 
PNS models, Fig. 5. The low - $\rho_{\rm centr}$ segments of the 
$M - \rho_{\rm centr}$ plots for the PNS are dramatically 
different from that for the cold NS models. The relevance of this 
features for the stability of lower-mass PNS will be discussed in 
Sect. 7.  
%
%%%%%%%%%%%%%%%%%%%%%%%%%%%%
\section{Characteristic timescales}
The EOS of PNS is evolving with time, due mainly to the 
deleptonization process, which  changes the composition of the
matter, and also due to changes of the internal temperature of the
star. However, these changes occur on the timescales $\tau_{\rm
evol}\sim $1-10~s (see below), which are
three or more orders of magnitude longer than the dynamical
timescale, governing the readjustment of pressure and gravity
forces. This dynamical timescale  $\tau_{\rm dyn}\sim 1~$ms
corresponds also  to the  characteristic periods of the PNS
pulsations. In view of this, we are
able to decouple the PNS evolution from its dynamics, with a well defined 
EOS of the PNS matter. 

The evolution of the PNS interior results from cooling,
deleptonization, and  dissipative transport processes, leading
to heating. Deleptonization of PNS is due to diffusion of
$\nu_e$, driven by the gradient of their chemical potential, and
occurs on a timescale  $\tau_{\rm delept}\sim$ few seconds
(Prakash et al. 1997). 
(Both $\tau_{\rm delept}$ and $\tau_{\rm cool}$ can actually 
be shorter, due to the presence of convection in some layers 
of a PNS).   
The PNS cooling is due to neutrino losses from the
neutrinosphere; for the neutrino-opaque interior,
 the characteristic timescale of cooling  
 $\tau_{\rm cool}$ is of the order of tens
of seconds (Sawyer \& Soni 1979). Notice, that deleptonization
is accompanied by a significant heating of neutrino-opaque core.
 Highly degenerate $\nu_e$ diffuse out from the neutrino-opaque
core, and due to $E_{\nu_e}\gg T$ they deposit most of 
 their energy in the matter, which in view of 
$\tau_{\rm cool}\gg \tau_{\rm delept}$ corresponds to the net
heating. 

Radial pulsations of PNS are damped due to dissipative
processes, resulting from the weak interactions involving
nucleons and leptons. The dissipative effects can be represented
in the form of a bulk viscosity of hot dense matter (Sawyer 1980).
Let us consider first the case of hot interior with a significant
trapped lepton number. For $Y_l=0.3$  Sawyer (1980) gets 
$\tau_{\rm damp}(Y_l=0.3)\sim 2000~(T/10~{\rm MeV})^2~$s. For
deleptonized hot interior the characteristic timescale is
somewhat shorter, 
$\tau_{\rm damp}(Y_\nu=0)\sim 30~(T/10~{\rm MeV})^2$~s (Sawyer
1980). Still, in both cases we get $\tau_{\rm damp}\widetilde> 
10^4\tau_{\rm dyn}$, so that one can safely neglect damping when
calculating the eigenfrequencies of radial pulsations of PNS. 

Summarizing, the estimates of the evolutionary and dissipative
timescales indicate, that radial pulsations of the hot interior
of PNS can be treated as adiabatic, and can be studied using a
well defined EOS of dense hot matter, corresponding to a given
stage of evolution of a PNS.  

Of course, all these remarks are valid only for the hot interior of
PNS (i.e., the region below the neutrinosphere). However, the
outer envelope contains less than $10^{-3}$ of the mass of PNS,
and its influence on the eigenfrequencies of radial pulsations
is negligible.  
%
%%%%%%%%%%%%%%%%%%%%%%%%%%%%%%%%%%%%%%%%%%%%%%%%%%%%%%%%%
\section{Linear adiabatic radial pulsations of protoneutron stars}
Consider an idealized static configuration of a PNS and assume 
it is spherically symmetric. Using 
the notation of Landau \& Lifshitz (1975), we write the 
 metric for such a configuration as
%%%%%%%%%%%%%%%%%%%%%%%%%%%%%%%%%%%%%%%%%%%%%%%
\begin{equation}
{\rm d}s^2=
{\rm e}^\nu c^2{\rm d}t^2
-{\rm e}^\lambda {\rm d}r^2
-r^2({\rm d}\theta^2+\sin^2\theta {\rm d}\phi ^2)~,
\label{eq:metric}
\end{equation}
%%%%%%%%%%%%%%%%%%%%%%%%%%%%%%%%%%%%%%%%%%%%%%%%
where $\lambda$ and $\nu$ are functions of $r$. 

The  hydrostatic equilibrium  of the static PNS is described 
by the
Tolman-Oppenheimer-Volkoff (TOV) equations (Tolman 1939, 
Oppenheimer \&
Volkoff 1939)
%%%%%%%%%%%%%%%%%%%%%%%%%%%%%%%%%%%%%%%%%%%%%
\begin{equation}
\label{a}
{{\rm d}P\over {\rm d}r}=
-{Gm\rho \over r^2 (1-{2Gm\over rc^2})}
\left(1+{P\over \rho c^2 }\right)
\left(1+{4\pi Pr^3\over m c^2}\right)~,
%\left(1-{2Gm\over rc^2}\right)^{-1}~  
\end{equation}
%%%%%%%%%%%%%%

\begin{equation}
\label{b}
{{\rm d}m\over {\rm d}r}=4\pi r^2\rho~,  
\end{equation}
%%%%%%%%%%%%%%%%
\begin{equation}
\label{c}
{{\rm d}\nu \over {\rm d}r}=-{2\over (P+\rho c^2 )}
{{\rm d}P\over {\rm d}r}~, 
\end{equation}
%%%%%%%%%%%%%%%%%%%%%%%%%%%%%%%%%%%%%%%%%%%%%%%
where $m$ is the mass contained within radius $r$. 
Since our EOS of PNS can be always written in the 
one-parameter form 
 $P=P(\rho)$, the TOV equations 
 can be numerically integrated for a given 
central density $\rho_{\rm centr}$,  yielding  the 
stellar radius, $R,$ 
and the total gravitational mass, 
$M=m(R)$,  of the star.

The equations governing infinitesimal radial adiabatic 
 stellar pulsations in general
relativity were derived by Chandrasekhar (1964), 
and were rewritten by Chanmugan (1977) in a form, which 
turns out to be particularly suitable for numerical 
applications. 
 Two important quantities, describing  pulsations, are:   
 the relative radial displacement, $\xi = \Delta r/r$, 
 where $\Delta r$ is the radial displacement of a matter element,  
and $\Delta P$  - the corresponding
  Lagrangian perturbation of the pressure. These two quantities 
are determined from a system of two ordinary differential 
equations, which we rewrite as 
%%%%%%%%%%%%%%%%%%%%%%
\begin{eqnarray}
{{\rm d}\xi \over {\rm d}r}&=&
-{1\over r}\left(3\xi+{\Delta P\over \Gamma P}\right)-
{{\rm d}P\over {\rm d}r}{\xi\over (P+\rho c^2)}~, \label{d}\\
{{\rm d}\Delta P \over {\rm d}r}&=&
\xi\left\{ 
{\omega^2 \over c^2}
{\rm e}^{\lambda-\nu}\left( P+\rho c^2\right)r
-4{{\rm d}P \over {\rm d}r}\right\}\nonumber\\
%\mbox{}
&+&\xi\left\{
 \left({{\rm d}P \over {\rm d}r}\right)^2 
{r \over(P+\rho c^2)}
-{8\pi G \over c^4}{\rm e}^\lambda (P+\rho c^2)Pr\right\}\nonumber\\
%\mbox{}
&+&{\Delta P} \left\{{{\rm d}P \over {\rm d}r}{1 \over
(P+\rho c^2)}-{4\pi G \over c^4} 
(P+\rho c^2)r{\rm e}^\lambda 
\right\}~, 
\label{e}
\end{eqnarray}
%%%%%%%%%%%%%%%%%%%%%%%%%%%%%%%%%
where $\Gamma$ is a relativistic adiabatic index 
(see Section 6),
$\omega$ is the eigenfrequency and the 
quantities $\xi$ and $\Delta P$ are
assumed to have a harmonic time dependence 
$\propto e^{i\omega t}$. 
Our Eq. (\ref{e}) has been obtained from a second-order 
pulsation equation of Chandrasekhar (1964) [his Eq. (59)], 
using his Eq. (35) and Eq.(36). While Eq. (\ref{d}) coincides with 
 Eq.(18) of Chanmugan (1977), 
our Eq. (\ref{e}) - in contrast to his 
second equation [Eq.(19) of Chanmugan 1977] 
 - does not show a singularity at the 
stellar surface. 
  Another important advantage of  our system of pulsation equations, 
 Eq. (\ref{d},~\ref{e}),  
 stems from the fact, that they do not involve  
 any derivatives of the adiabatic index, 
$\Gamma.$ In view of the fact, that we used tabulated
forms of the EOS and of $\Gamma$,  this enabled us to 
reach a very high
precision in solving the eigenvalue problem.  

To solve equations
(\ref{d}) and (\ref{e}) one needs two boundary conditions.  
 The condition of regularity at $r=0$ requires,  that for 
$r\rightarrow 0$ the
coefficient of the $1/r$-term  in Eq. (\ref{d}) must vanish,
%%%%%%%%%%%%%%%%%%%%%%%%%
\begin{equation}
\label{f}
\left(\Delta P\right)_{\rm center}=
-3\left(\xi \Gamma P\right)_{\rm center}~.
\end{equation}
%%%%%%%%%%%%%%%%%%%%%%%%%
Our normalization of eigenfunctions corresponds to  $\xi(0)=1$. 
 The surface of the star is determined by 
 the condition that for  $r\rightarrow R$,  
one has $P\rightarrow 0$. This implies 
%%%%%%%%%%%%%%%%%%%%%%
\begin{equation}
\label{g}
\left(\Delta P\right)_{\rm surface} =0
\end{equation}
%%%%%%%%%%%%%%%%%%%%%

The problem of solving Eqs. (\ref{d}), (\ref{e}), can be reduced 
to a second order (in $\xi$), 
linear radial wave 
equation, of the Sturm-Liouville type. The quantity $\omega^2$ is 
the eigenvalue of the Sturm-Liouville problem with boundary conditions 
given by Eqs. (\ref{f}), (\ref{g}).  
For a given static model of a PNS, we get a set of the 
eigenvalues  $\omega_0^2 < \omega_1^2 
<...<\omega_{\rm n}^2<...,$  with corresponding 
eigenfunctions $\xi_0,\xi_1,...,
\xi_{\rm n},...,$ where the eigenfunction $\xi_{\rm n}$ has 
$n$ nodes  within the star, $0\le r \le  R$ (see, e.g., Cox 1980). 

A given static model is stable with respect to small, radial, adiabatic 
perturbations (pulsations),  if $\omega_{\rm n}^2 > 0$ for all $n$. 
The
 configuration is marginally stable, if
 the lowest eigenfrequency $\omega_0 = 0$. 
%%%%%%%%%%%%%%%%%%%%%%%%%%%%%%%%%%%%%%%%%%%%%%%%%
\section{Adiabatic indices}
The adiabatic index within the star, $\Gamma$, plays  a central
role for both the linear radial oscillations, and for the
stability of PNS. In what follows, we will restrict ourselves to
the case of the neutrino-opaque interior; the layer above the
neutrinosphere contains only about $10^{-3}$ of the total mass,
and does not influence neither the  spectrum of radial
oscillations, nor the stability of PNS. 

Under physical conditions prevailing within the hot interior of
a PNS, the perturbation of a local nucleon density, $\delta n$,
of a matter element during radial oscillations, takes place at
constant entropy per baryon, $s$, and constant electron lepton
number per baryon, $Y_l$. Due to high density and temperature,
all constituents of matter can be considered as being 
in thermodynamic equilibrium  (the timescale of reactions
leading to thermodynamic equilibrium, $\tau_{\rm react}$, is
much shorter, then the pulsation timescale, 
$\tau_{\rm puls}\sim \tau_{\rm dyn}$).
The 
adiabatic index, governing linear perturbation of the pressure
within the star under these conditions, will be denoted by
$\Gamma_{\rm a}$. It will be given by
%%%%%%%%%%%%%%%
\begin{equation}
\Gamma_{\rm a} \equiv  
      {n\over P}\left({{\rm d}P\over {\rm d}n}\right)_{s,Y_l}~,
\label{Gamma_a}
\end{equation}
%%%%%%%%%%%%%
where the derivative is to be calculated at fixed $s$ and $Y_l$.
 It should be stressed, that the calculation of $\Gamma_{\rm a}$
requires more than just the knowledge of the EOS of the PNS
interior. Namely, one needs to know the {\rm perturbed} EOS,
under the constraint  of constant $s$ and $Y_l$, and assuming
the thermodynamic equilibrium of matter constituents. 

In the case of dense  matter one can  define several
$\Gamma$'s, which are physically (and numerically) 
different from $\Gamma_{\rm a}$. The EOS of dense hot matter
within a static PNS  
yields the quantity
%%%%%%%%%%%
\begin{equation}
\Gamma_{\rm EOS} \equiv  
      {n\over P}\left({{\rm d}P\over {\rm d}n}\right)_{\rm star}~.
\label{Gamma_EOS}
\end{equation}
%%%%%%%%%%%%%%%%%%%%%%%
Strictly speaking $\Gamma_{\rm EOS}$ is not  an ``adiabatic'' index
unless the entropy $s$ is constant throughout a star.

In the case of a sufficiently cold matter, the reactions between
matter 
constituents are so slow, that the matter composition remains
fixed (frozen) during perturbations, because 
$\tau_{\rm react}\gg \tau_{\rm dyn}$. Also, neutrinos do not
contribute then to the thermodynamic quantities of the matter. 
In such a case, the
appropriate adiabatic index would read
%%%%%%%%%%%%%%%%%%%%%%%%%%%%%
\begin{equation}
\Gamma_{\rm frozen} \equiv  
      {n\over P}\left({{\rm d}P\over {\rm d}n}\right)_{s,Y_e}~,
\label{Gamma_frozen}
\end{equation}
%%%%%%%%%%%%%%%%%%%%%%%%%%%%%
where $s,Y_e$ correspond to the equilibrium model. 
The quantity $\Gamma_{\rm frozen}$ is relevant for radial
pulsations of standard, cold neutron stars, where the difference
between $\Gamma_{\rm frozen}$ and $\Gamma_{\rm EOS}$ has
interesting consequences for the stability of the cold  NS models in
the vicinity of $M_{\rm max}$ (Gourgoulhon et al. 1995). 

%%%%%%%%%%%%%%%%%
\begin{figure}     %6
\begin{center}
\leavevmode
\epsfxsize=8.5cm \epsfbox{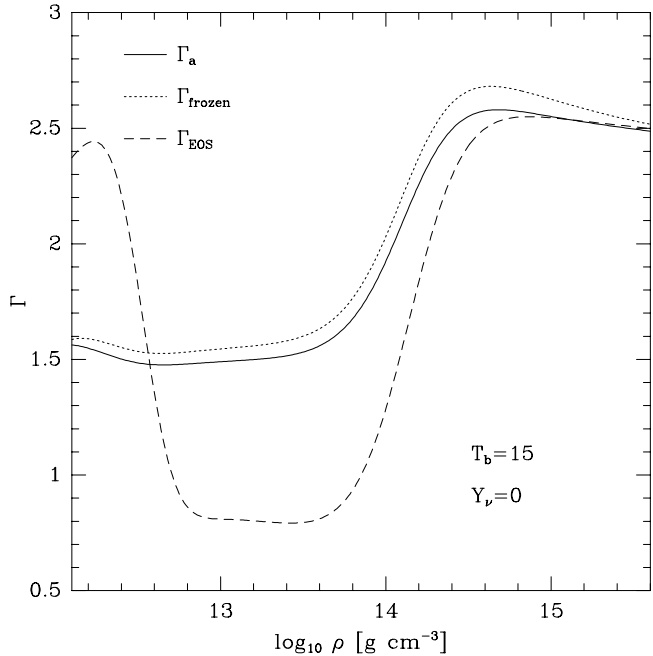}
\end{center}
\caption[]{
Parameters $\Gamma_{\rm a}$, $\Gamma_{\rm EOS}$, and
$\Gamma_{\rm frozen}$, versus matter density, for the isothermal
hot interior with $T_{\rm b}=15~{\rm MeV}$ and with no trapped lepton
number.    The outer edge of the hot
core has been located at 
%$n_\nu=2\times 10^{-3}~{\rm fm^{-3}}$,
$\rho_\nu=3.5\times 10^{12}~{\rm g\,cm^{-3}}$,
and the rapid changes in the $\Gamma$'s below $n_\nu$ results
from the temperature drop in the neutrinosphere of the PNS.
}
\label{fig6}
\end{figure}
%%%%%%%%%%%%%%%%%%%%%%%%

%%%%%%%%%%%%%%%%%
\begin{figure}     %7
\begin{center}
\leavevmode
\epsfxsize=8.5cm \epsfbox{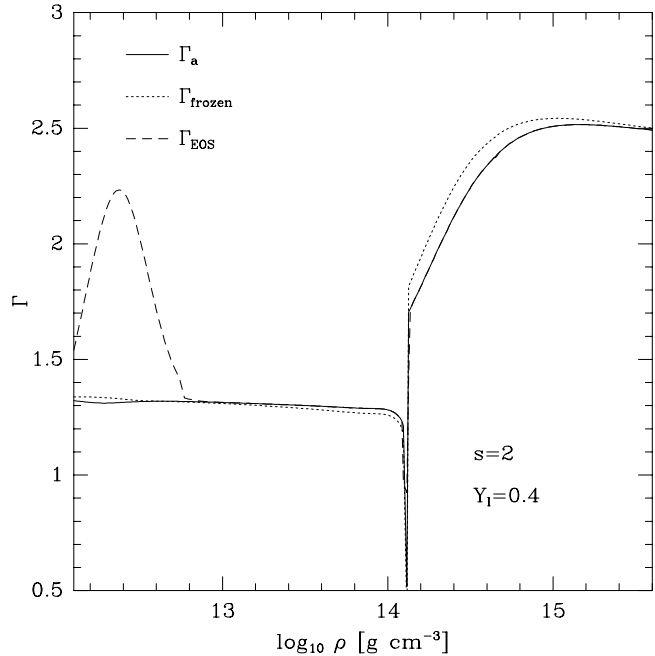}
\end{center}
\caption[]{
Parameters $\Gamma_{\rm a}$, $\Gamma_{\rm EOS}$, and
$\Gamma_{\rm frozen}$, versus matter density, for the isentropic
 hot interior with $s=2$ and with trapped lepton fraction 
 $Y_l=0.4$.  The outer edge of the hot
core has been located at 
%$n_\nu=2\times 10^{-3}~{\rm fm^{-3}}$,
$\rho_\nu=3.5\times 10^{12}~{\rm g\,cm^{-3}}$,
and the rapid changes in the $\Gamma$'s below $n_\nu$ results
from the temperature drop in the neutrinosphere of the PNS. The
very steep drop in $\Gamma$'s  above $10^{14}~{\rm g\,cm^{-3}}$ results
from the phase transition from matter with nuclei to bulk
(homogeneous) dense matter. 
}
\label{fig7}
\end{figure}
%%%%%%%%%%%%%%%%%%%%%%%

The dependence of various $\Gamma$'s on the matter density in
the PNS interior, for two models of the hot neutrino-opaque core
of PNS, is displayed in Fig. 5, 6. Above nuclear density (i.e.,
for bulk, homogeneous  matter) $\Gamma_{\rm frozen}$ is the
highest of all $\Gamma$'s; freezing of lepton composition
stiffens the matter (c.f. $\S$81 of Landau \& Lifshitz (1987)).
In the case of the isothermal core, Fig. 6, $\Gamma_{\rm EOS}$
is significantly lower than $\Gamma_{\rm a}$ and $\Gamma_{\rm
frozen}$, for subnuclear densities  above $n_\nu$  
 (this results from the fact of a significant  density dependence
of the  entropy per baryon, in this density interval). 
 For isothermal models  $\Gamma_{\rm EOS}\equiv \Gamma_{\rm star} 
< \Gamma_{\rm a}$, except for the region close to and below 
the neutrinosphere: for $\rho>10^{13}~{\rm g~cm^{-3}}$ our 
 isothermal PNS models are convectively stable.  
 The differences between various $\Gamma$'s 
differences are much smaller in the case of the isentropic core,
where fixing $s$ throughout the star and within the EOS makes
$\Gamma_{\rm EOS}$ and $\Gamma_{\rm a}$ undistinguishable,
except near the neutrinosphere, where a dramatic  drop in $s$
and  in $T$ produces a strong deviation of 
$\Gamma_{\rm EOS}$ 
from both $\Gamma_{\rm a}$ and $\Gamma_{\rm frozen}$.   
  The differences between $\Gamma_{\rm frozen}$ and $\Gamma_{\rm a}$
are mainly due to the changes of $Y_e$ throughout the star.
 The quantity $\Gamma_{\rm frozen}$ corresponds to adiabatic 
pulsations keeping $Y_e$ 
fixed, while $\Gamma_{\rm a}$ was calculated assuming  beta equilibrium 
during adiabatic oscillations at fixed $Y_l$.  
It should be
stressed, that pulsational properties of PNS are determined
essentially by the values of the relevant $\Gamma$ well above
the density $n_\nu$. The outer layer with $n<n_\nu$ contains a
very small fraction of stellar mass, 
and therefore rapid variations of $\Gamma$
close to $n_\nu$ have no effect on the global dynamics of PNS. 
%
%%%%%%%%%%%%%%%%%%%%%%%%%%%%%%%%%%%%%%%%%%%%%%%%%%%%%%%%%
\section{Eigenfrequencies and instabilities} 
The effects of high temperature and those of the trapped
neutrinos  influence both the EOS  and the adiabatic index of
the interior of PNS. Our calculations  show a rather strong
influence of these effects on the eigenfunctions and  the 
eigenfrequencies of radial pulsations of PNS. Also, these
effects imply significant differences, especially within 
some (observationally interesting) interval of stellar masses,
between pulsational properties of PNS and cold NS. 

%%%%%%%%%%%%%%%%%
\begin{figure}     %8
\begin{center}
\leavevmode
\epsfxsize=8.5cm \epsfbox{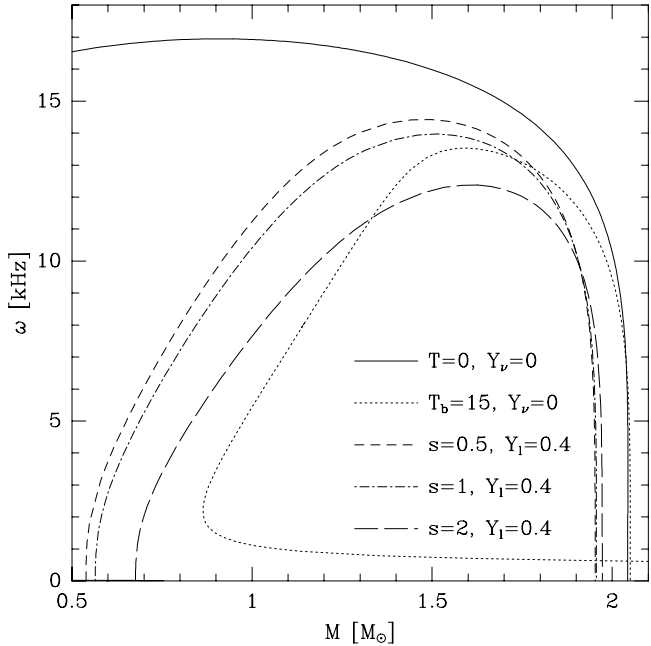}
\end{center}
\caption[]{
The eigenfrequencies of the fundamental mode, $n=0$,     
versus stellar mass for considered models of dense matter.
The nearly horizontal part of the curve for 
$T_{\rm b}=15$~MeV model corresponds
to the configurations with central density below the value 
corresponding to the minimum mass of PNS.
 For the sake of comparison we also show corresponding curve
 for cold neutron stars. 
}
\label{fig8}
\end{figure}
%%%%%%%%%%%%%%%%%%%%%%%

In Fig. 8  we plotted the eigenfrequency of the 
fundamental mode of the PNS pulsations, versus  
stellar mass, for several models of the EOS of the hot 
interior. 
 For the sake of comparison, we have  presented also the
corresponding plot for  cold neutron stars. 
 The effect of finite entropy and of trapped
lepton number on $\omega_0$ depends rather strongly 
on the stellar mass. The
relativistic instability takes place very close to $M_{\rm
max}$, so that the ``classical stability criterion''
(protoneutron star models to the left of the maximum in the $M -
R$ plot, Fig. 4, are secularly unstable with respect to the
$n=0$ oscillations) is valid to a very good approximation.
Generally, entropy and trapped lepton number soften the $n=0$
mode with respect to the $T=0$ models.  This softening is
particularly strong for lower stellar mass. In the case of
$s=2$, $Y_l=0.4$ PNS, the fundamental mode becomes secularly
unstable at $M=0.676~M_\odot$, very close to the
minimum of the $M - R$ curve, Fig. 4. The classical stability
criterion gives therefore a rather precise location of the
 instability point  for the {\it isentropic} PNS: the models to
the right of the minimum in the $M - R$ curve are 
unstable with  respect to the radial pulsations in the
fundamental mode. However, the value of $M_{\rm min}$ is
dramatically larger than that for the {\it cold} ($T=0$) neutron
stars. We have $M_{\rm min}[s=2,Y_l=0.4]=0.675~M_\odot$, 
while $M_{\rm min}[T=0]=0.054~M_\odot$.
%%%%%%%%%%%%%%%%%
\begin{figure}     %9
\begin{center}
\leavevmode
\epsfxsize=8.5cm \epsfbox{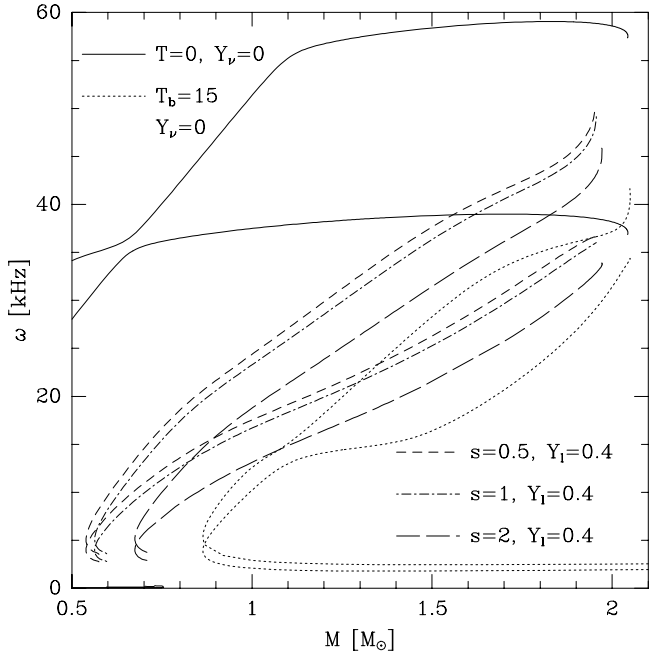}
\end{center}
\caption[]{
The eigenfrequencies of the first and second overtone, $n=1,2$,     
versus stellar mass for considered models of dense matter.
For the sake of comparison we also show corresponding curves
 for cold neutron stars. 
}
\label{fig9}
\end{figure}
%%%%%%%%%%%%%%%%%%%%%%%

In Fig. 9 we plotted the eigenfrequencies of the 
 $n=1,~2$  radial modes versus stellar mass, for the isentropic 
models of the hot PNS cores. For the sake of comparison, 
we have shown also the corresponding plots for cold NS. 
 The differences
between the NS and the PNS  $n=1,~2$ eigenfrequencies are very large,
except for a narrow interval of masses in the vicity of the
maximum mass. These differences reflect the drastic differences
in the structure of the outer layers of the PNS and the NS
models. Generally, PNS are much softer with respect to the
$n=1,~2$ radial modes. 
 Different
structure of the outer layers of stellar models with isentropic
and isothermal cores, implies strong differences in the
spectra of the lowest three modes of radial pulsations. 
Clearly, the $M<1~M_\odot$ isothermal models are much ``softer''
with respects to all considered modes of radial pulsations, than
the  isentropic ones; this is due to the fact that they are
significantly less compact  (inflated by the thermal effects)
than their isentropic counterparts.  
 The characteristic "turning points" and the intermodal 
intersections 
 for the PNS models on the low-mass side of Fig. 9,  
  are due to the existence of minimum masses. 
 However, it should be stressed that the frequency intersection 
points,  seen in the case of the case of the $n=1$ and the $n=2$ 
eigenfrequencies, do not correspond to the same configuration 
of the PNS  (while having the same mass, these configurations 
 have different central densities, see Figures 10, 11). 

The situation is less complicated 
%\jlz and the thermal effect on the pulsational spectrum looks less dramatic
in the case of the plots 
of the eigenfrequencies versus the central density
of  the stellar models,  $\rho_{\rm centr}$,
presented  in Figures 10 and 11. 
 At lower central densities the values of
$\omega_{\rm n}$ decrease monotonically, but very slowly, with
decreasing central density.   
%%%%%%%%%%%%%%%%%
\begin{figure}     %10
\begin{center}
\leavevmode
\epsfxsize=8.5cm \epsfbox{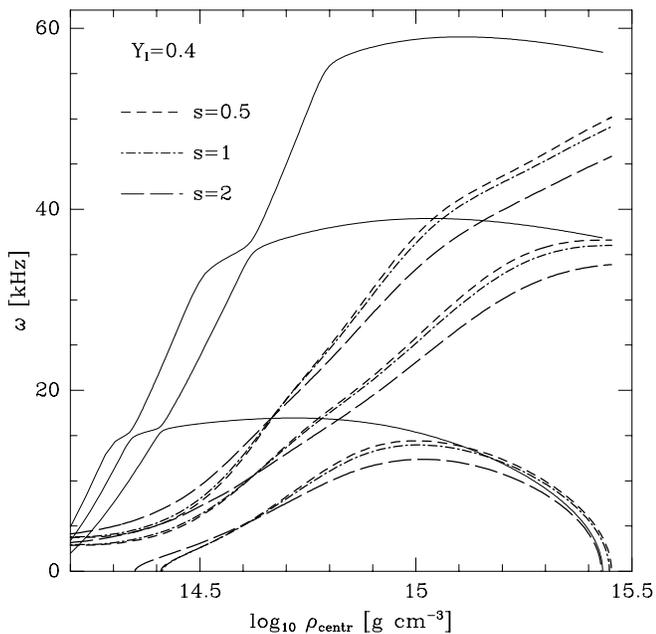}
\end{center}
\caption[]{
The eigenfrequencies of three lowest radial modes, 
$n=0,~1,~2$,   for hot isentropic protoneutron stars with  
trapped lepton fraction $Y_l=0.4$, versus central density. 
 For the sake of comparison we also show corresponding curves
 for cold neutron stars (thin solid lines). 
}
\label{fig10}
\end{figure}
%%%%%%%%%%%%%%%%%%%%%%%
%%%%%%%%%%%%%%%%%
\begin{figure}     %11
\begin{center}
\leavevmode
\epsfxsize=8.5cm \epsfbox{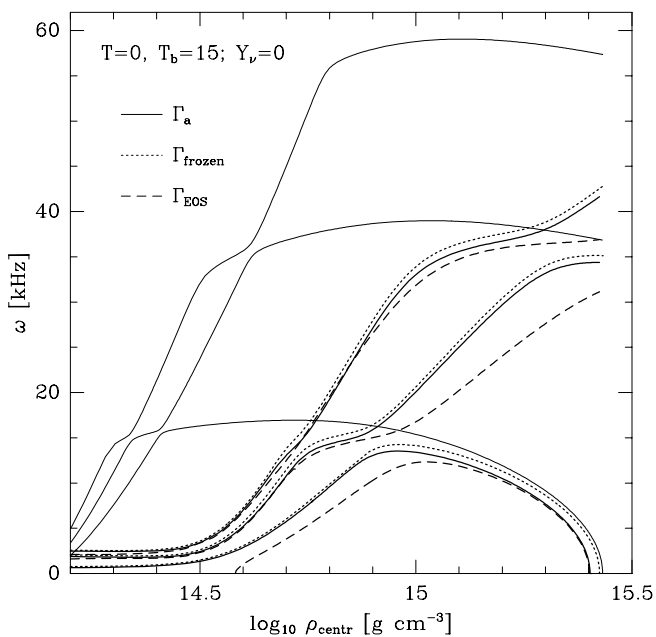}
\end{center}
\caption[]{
The eigenfrequencies of three lowest radial modes, 
, $n=0,~1,~2$,   for protoneutron stars with  hot isothermal
core  with $T_{\rm b}=15~{\rm MeV}$,  versus central density 
For the sake of comparison we also show corresponding curves 
 for cold neutron stars (thin solid lines). 
}
\label{fig11}
\end{figure}
%%%%%%%%%%%%%%%%%%%%%%%
%

 The effects of temperature and
  of the differences in adiabatic indices $\Gamma$  
are particularly strong
in the case of the isothermal hot interior, with
$T=T_\infty (g_{00})^{-1/2}$. Our results for the first three radial
modes ($n=0,~1,~2$), for $T_{\rm b}=15~$MeV,  are given in Fig. 11, 
where we show also for comparison 
results obtained for cold NS. 
  Three curves presented in Fig. 11 correspond to three different
choices of $\Gamma$, governing the changes of pressure vs. density during
pulsations. Although only results for $\Gamma=\Gamma_{\rm a}$ are physical,
the difference between curves visualizes the role of the proper treatment
of chemical equilibrium and adiabacity when the PNS star is oscillating.
$\Gamma_{\rm EOS}$ is determined by the equation of state of the matter
 within a {\it static} model of PNS. 
 As one can see from  Fig. 11, 
  the use of $\Gamma_{\rm EOS}$ leads usually to a significant 
underestimating  of the 
eigenfrequencies of oscillation modes. A notable  
 exception  from this behavior is that of the fundamental mode near  
 the maximum mass of PNS, for which results obtained using 
$\Gamma_{\rm EOS}$ are very close to those obtained using 
 $\Gamma_{\rm a}$.  
 Consequently,  the classical static stability
criterion at $M_{\rm max}$ works rather well also for  the {\it
isothermal} PNS.
 
However, 
 the difference between $\omega_{0}^{\rm (EOS)}$ and  
actual  $\omega_0$
increases with decreasing $\rho_{\rm centr}$ and
 the static stability criterion at
$M_{\rm min}$ does not hold. We have $M_{\rm min}[T_{\rm b}=15~{\rm
MeV}]=0.86~M_\odot$
 but configurations in the neighbourhood of the
 mass minimum turn out to be stable with respect to the fundamental
mode of the radial pulsations, calculated using the {\it actual} 
(physical) values of $\Gamma=\Gamma_{\rm a}$. 

In Fig. 11 we can see an interesting phenomenon --- 
abrupt, nearly stepwise
changes in frequencies of the consecutive  
(neighbouring) oscillation modes, 
especially pronounced  
for cold neutron stars, 
 but clearly visible also in the case of
$T_{\rm b}=15$~MeV, $Y_\nu=0$ models of PNS.
 This is the ``avoided crossing'' phenomenon
of the {\it radial} modes of neutron stars. 
This effect is known from the analysis of the
{\it nonradial} oscillations in ``ordinary''  stars (Aizenman, Smeyers 
and Weigert (1977), Christensen--Dalsgaard (1980)) and 
has been recently
considered by Lee and Strohmayer (1996) for {\it nonradial} oscillations of
rotating neutron stars. 
 Our  preliminary studies of this effect indicate, 
 that the stepwise changes of $\omega_n$  are due to the change 
of the character of the standing-wave solution for the eigenproblem. 
Namely, at the "avoided crossing' point the solution changes from the standing wave localized mainly 
in the outer layer of the star, to that localized predominantly  in the 
central core.  
This topic will be discussed in separate paper (Gondek et al. 1996). 

%%%%%%%%%%%%%%%%%%%%%%%%%%%%%%%%%%%%%%%%%%%%%%%%%%%%%%%%%
\section{Discussion and conclusions}
The birth of protoneutron stars is a dynamical process, and a 
newborn protoneutron star is expected to pulsate. 
These pulsations were  excited during star formation. In the
present paper we studied linear, radial pulsations of protoneutron
stars, under various assumptions concerning hot stellar core,
and for a large interval of stellar masses. 

The spectrum of the lowest modes of radial pulsations of
protoneutron stars is quite different from that of cold neutron
stars. Generally, protoneutron stars are significantly softer
with respect to the radial pulsations, than cold neutron stars,
and this difference increases for higher modes and for lower
stellar masses. These differences stem from different structure
of protoneutron stars, which in contrast to cold neutron stars  
have extended envelopes, inflated
by thermal and trapped neutrinos effects. 

The standard static criteria of stability of neutron star
models were derived under the assumption of cold, catalyzed
matter (Harrison et al. 1965). We have shown, that to a rather
good approximation, the configuration with maximum mass
separates stable configurations from the secularly
unstable ones (with respect to fundamental mode of small radial
pulsations). Therefore, despite thermal and neutrino trapping
effects, one can apply standard  ``maximum mass'' criterion to locate
the relativistic  instability of
protoneutron stars, both for the isentropic and the isothermal
hot, neutrino opaque cores. 

The role of thermal effects increases with decreasing stellar
mass. Static ``minimum mass'' criterion does apply with a rather
good precision to protoneutron stars with hot isentropic,
neutrino opaque cores. This is due to the fact, that the
perturbation itself conserves entropy (pulsations are
adiabatic), and equilibration of matter is sufficiently fast.
However, the ``minimum mass'' criterion does not apply to
protoneutron stars with hot isothermal cores. In both cases, the
minimum masses of protoneutron stars are rather large (for our
models of dense hot matter we obtained the values of
$0.675~M_\odot$ and $0.86~M_\odot$ for the isentropic 
($s=2$, $Y_l=0.4$) and the isothermal ($T_{\rm b}=15~{\rm MeV}$)
models, respectively. 

Our treatment of the thermal state of the  protoneutron star
interior should be considered as very crude. The temperature
profile might be  affected by  convection. Also, our method of
locating the neutrinosphere was very approximate. Clearly, the
treatment of thermal effects can be refined, but we do not think
this will significantly change our main results. 

Our calculations were performed for only one model of the
nucleon component of dense hot matter. The model was realistic,
and enabled us to treat in a unified way the whole interior
(core as well as  the envelope) of
the protoneutron star. However, in view of the uncertainties in
the EOS of dense matter at supranuclear densities, one should of
course study the whole range of theoretical possibilities, for a
broad set - from soft to stiff - of supranuclear, high
temperature EOS.  An example of such an investigation 
in the case of the {\it static} protoneutron stars, 
is the study of Prakash et al. (1997). 

In a hypothetical scenario, proposed by Brown and Bethe (1994),
a protoneutron star borns  as a hot, lepton rich object,
composed of nucleons and leptons. If the final state of cold
neutron star is characterized by a large electron fraction, due
to the appearance of a large amplitude $K^-$ - condensate 
(or
because a large fraction of baryons are negatively charged
hyperons), then the  cold equation of state of neutron star
matter is softer than that of hot, lepton rich protoneutron
star. In such a case, 
$M_{\rm bar,max}({\rm NS})<M_{\rm bar,max}({\rm PNS})$, 
and all protoneutron stars with baryon mass
exceeding  $M_{\rm bar,max}({\rm NS})$ will eventually collapse into black
holes. Clearly, in such a case the problem of stability
of the protoneutron star models in the vicinity of 
$M_{\rm bar,max}({\rm PNS})$ with respect to radial
pulsations   looses its significance. 

\begin{acknowledgements}
 We are very grateful to W. Dziembowski for introducing us into 
the topic of  the ``avoided crossing'' phenomena,  for his 
numerous helpful remarks, and for careful reading of the 
manuscript.    
This research was partially supported by the KBN grant No. P304
014 07  and  by the KBN grant No. 2P03D01211 for D. Gondek. 
 D. Gondek and P. Haensel were also supported by the program 
R{\'e}seau Formation Recherche of the French Minist{\`e}re 
de l'Enseignement Sup{\'e}rieure et de la Recherche. 
\end{acknowledgements}
\par

\end{document}